\documentstyle[11pt,graphicx]{article}                
\def\st{Stueckelberg ~}

\def\rmuu{\gamma^{\mu}}
\def\rmud{\gamma_{\mu}}
\def\PL{{1-\gamma_5\over 2}}
\def\PR{{1+\gamma_5\over 2}}
\def\sinW2{\sin^2\theta_W}
\def\AEM{\alpha_{EM}}
\def\mul{M_{\tilde{u} L}^2}
\def\mur{M_{\tilde{u} R}^2}
\def\mdl{M_{\tilde{d} L}^2}
\def\mdr{M_{\tilde{d} R}^2}
\def\mz2{M_{z}^2}
\def\c2b{\cos 2\beta}
\def\au{A_u}
\def\ad{A_d}
\def\cob{\cot \beta}
\def\v#1{v_#1}
\def\tb{\tan\beta}
\def\epem{$e^+e^-$}
\def\KK{$K^0$-$\overline{K^0}$}
\def\wi{\omega_i}
\def\xj{\chi_j}
\def\Wmu{W_\mu}
\def\Wnu{W_\nu}
\def\m#1{{\tilde m}_#1}
\def\mH{m_H}
\def\mw#1{{\tilde m}_{\omega #1}}
\def\mx#1{{\tilde m}_{\chi^{0}_#1}}
\def\mc#1{{\tilde m}_{\chi^{+}_#1}}
\def\mwi{{\tilde m}_{\omega i}}
\def\mxi{{\tilde m}_{\chi^{0}_i}}
\def\mci{{\tilde m}_{\chi^{+}_i}}

\def\ch{{\tilde\chi^{+}_1}}
\def\c2{{\tilde\chi^{+}_2}}

\def\tt{{\tilde\theta}}

\def\tp{{\tilde\phi}}

\def\mz{M_z}
\def\sw{\sin\theta_W}
\def\cw{\cos\theta_W}
\def\cb{\cos\beta}
\def\sb{\sin\beta}
\def\rwi{r_{\omega i}}
\def\rxj{r_{\chi j}}
\def\rfp{r_f'}
\def\Kik{K_{ik}}
\def\Fq2{F_{2}(q^2)}
\def\f{\({\cal F}\)}
\def\d1{{\f(\tilde c;\tilde s;\tilde W)+ \f(\tilde c;\tilde \mu;\tilde W)}}
\def\tw{\tan\theta_W}
\def\sec2w{sec^2\theta_W}
\begin{document}
\baselineskip 18pt
\def\today{\ifcase\month\or
 January\or February\or March\or April\or May\or June\or
 July\or August\or September\or October\or November\or December\fi
 \space\number\day, \number\year}
\def\thebibliography#1{\section*{References\markboth
 {References}{References}}\list
 {[\arabic{enumi}]}{\settowidth\labelwidth{[#1]}
 \leftmargin\labelwidth
 \advance\leftmargin\labelsep
 \usecounter{enumi}}
 \def\newblock{\hskip .11em plus .33em minus .07em}
 \sloppy
 \sfcode`\.=1000\relax}
\let\endthebibliography=\endlist
\def\lsim{\ ^<\llap{$_\sim$}\ }
\def\gsim{\ ^>\llap{$_\sim$}\ }
\def\r2{\sqrt 2}
\def\beq{\begin{equation}}
\def\eeq{\end{equation}}
\def\beqn{\begin{eqnarray}}
\def\eeqn{\end{eqnarray}}
\def\rmuu{\gamma^{\mu}}
\def\rmud{\gamma_{\mu}}
\def\PL{{1-\gamma_5\over 2}}
\def\PR{{1+\gamma_5\over 2}}
\def\sinW2{\sin^2\theta_W}
\def\AEM{\alpha_{EM}}
\def\mul{M_{\tilde{u} L}^2}
\def\mur{M_{\tilde{u} R}^2}
\def\mdl{M_{\tilde{d} L}^2}
\def\mdr{M_{\tilde{d} R}^2}
\def\mz2{M_{z}^2}
\def\c2b{\cos 2\beta}
\def\au{A_u}         
\def\ad{A_d}
\def\cob{\cot \beta}
\def\v#1{v_#1}
\def\tb{\tan\beta}
\def\epem{$e^+e^-$}
\def\KK{$K^0$-$\bar{K^0}$}
\def\wi{\omega_i}
\def\xj{\chi_j}
\def\Wmu{W_\mu}
\def\Wnu{W_\nu}
\def\m#1{{\tilde m}_#1}
\def\mH{m_H}
\def\mw#1{{\tilde m}_{\omega #1}}
\def\mx#1{{\tilde m}_{\chi^{0}_#1}}
\def\mc#1{{\tilde m}_{\chi^{+}_#1}}
\def\mwi{{\tilde m}_{\omega i}}
\def\mxi{{\tilde m}_{\chi^{0}_i}}
\def\mci{{\tilde m}_{\chi^{+}_i}}
\def\mz{M_z}
\def\sw{\sin\theta_W}
\def\cw{\cos\theta_W}
\def\cb{\cos\beta}
\def\sb{\sin\beta}
\def\rwi{r_{\omega i}}
\def\rxj{r_{\chi j}}
\def\rfp{r_f'}
\def\Kik{K_{ik}}
\def\Fq2{F_{2}(q^2)}
\def\f{\({\cal F}\)}
\def\d1{{\f(\tilde c;\tilde s;\tilde W)+ \f(\tilde c;\tilde \mu;\tilde W)}}
\def\tw{\tan\theta_W}
\def\sec2w{sec^2\theta_W}
\def\ch{{\tilde\chi^{+}_1}}
\def\c2{{\tilde\chi^{+}_2}}

\def\tt{{\tilde\theta}}

\def\tp{{\tilde\phi}}

\def\mz{M_z}
\def\sw{\sin\theta_W}
\def\cw{\cos\theta_W}
\def\cb{\cos\beta}
\def\sb{\sin\beta}
\def\rwi{r_{\omega i}}
\def\rxj{r_{\chi j}}
\def\rfp{r_f'}
\def\Kik{K_{ik}}
\def\Fq2{F_{2}(q^2)}
\def\f{\({\cal F}\)}
\def\d1{{\f(\tilde c;\tilde s;\tilde W)+ \f(\tilde c;\tilde \mu;\tilde W)}}
\def\tw{\tan\theta_W}
\def\sec2w{sec^2\theta_W}

\def \cha{\widetilde{\chi}^{\pm}_1}
\def \chb{\widetilde{\chi}^{\pm}_2}

\def \na{\widetilde{\chi}^{0}_1}
\def \nb{\widetilde{\chi}^{0}_2}
\def \nc{\widetilde{\chi}^{0}_3}
\def \nd{\widetilde{\chi}^{0}_4}

\def \g{\widetilde{g}}
\def \ql{\widetilde{q}_L}
\def \qr{\widetilde{q}_R}

\def \dl{\widetilde{d}_L}
\def \dr{\widetilde{d}_R}
\def \ul{\widetilde{u}_L}
\def \ur{\widetilde{u}_R}

\def \ccl{\widetilde{c}_L}
\def \ccr{\widetilde{c}_R}
\def \ssl{\widetilde{s}_L}
\def \ssr{\widetilde{s}_R}

\def \ta{\widetilde{t}_1}
\def \tb{\widetilde{t}_2}
\def \ba{\widetilde{b}_1}
\def \bb{\widetilde{b}_2}

\def \sta{\widetilde{\tau}_1}
\def \stb{\widetilde{\tau}_2}

\def \smr{\widetilde{\mu}_R}
\def \ser{\widetilde{e}_R}
\def \sml{\widetilde{\mu}_L}
\def \sel{\widetilde{e}_L}

\def \slr{\widetilde{l}_R}
\def \sll{\widetilde{l}_L}

\def \snl{\widetilde{\nu}_{\tau}}
\def \snm{\widetilde{\nu}_{\mu}}
\def \sne{\widetilde{\nu}_{e}}

\def \hc{H^{\pm}}

\def \lra{\longrightarrow}
\def\st{Stueckelberg~}
\def\s1{$s_{\alpha}$}
\def\s2{$s_{\gamma}$}
\def\s3{$s_{\delta}$}
\def\c1{$c_{\alpha}$}
\def\c2{$c_{\gamma}$}
\def\c3{$c_{\delta}$}
\def\y{Y_{\phi}}
\begin{titlepage}

\begin{center}

{~ {~ 
Predicted Signatures at the LHC from $U(1)$  Extensions of the Standard Model\footnote{Based on lectures at the 46th Course at the International School of Subnuclear Physics - Erice -Sicily: 29 August -7 September, 2008.}.
}}\\
\vskip 0.5 true cm
\vspace{2cm}
\renewcommand{\thefootnote}
{\fnsymbol{footnote}}
 Pran Nath  
\vskip 0.5 true cm
\end{center}

\noindent
{Department of Physics, Northeastern University,
Boston, MA 02115-5000, USA} \\
\vskip 1.0 true cm

\centerline{~ Abstract}
We discuss the $U(1)_X$ extensions of the standard model with focus on the Stueckelberg 
mechanism for mass growth for the extra $U(1)_X$ gauge boson.  The assumption of an 
axionic connector field which carries dual $U(1)$ quantum numbers,  i.e., quantum numbers
for the hypercharge $U(1)_Y$ and for the hidden sector gauge group $U(1)_X$, allows a 
non-trivial mixing between the  mass growth for the neutral gauge vector bosons in  the $SU(2)_L\times U(1)_Y$  
sector 
and the mass growth for the vector boson by  
 the \st mechanism in the $U(1)_X$ sector. This results in an
extra $Z'$ which can be very narrow, but 
still detectable at the Large Hadron Collider (LHC).   The $U(1)_X$ extension of the minimal supersymmetric
standard model is also considered and the role of the Fayet-Illiopoulos term in such an extension 
discussed.  The $U(1)_X$ extensions of the SM and of  the MSSM lead to new candidates 
for dark matter.

\medskip
\end{titlepage}

\noindent
{\em Introduction:}
 In lecture I we have discussed high scale models based on supersymmetry, local supersymmetry
 and  supergravity \cite{Nath:1975nj,Freedman:1976xh}. Specifically we focused on supergravity
 grand unified models\cite{msugra} and their experimental consequences, 
 their extensions and tests at the large Hadron Collider (LHC)\cite{fln}. 
 Here we discuss extensions of the
 standard model(SM) and of the minimal supersymmetric standard model(MSSM) to include
 extra $U(1)$ factors. Such 
 extra $U(1)$ factors appear in a variety of unified models: in grand unified models, i.e., in
string models  and in D brane models. Thus in $SO(10)$ and $E_6$ grand unification
one has 
\beqn
SO(10) \supset SU(5)\times U(1)_{\chi}; ~E_6\supset SO(10)\times U(1)_{\psi}
\eeqn  
Similarly D brane constructions naturally have many $U(1)$ factors since here on typically starts 
with a stack of n branes which has a $U(n)$ gauge symmetry. Since $U(n)\supset SU(n)\times U(1)$,
one has $U(1)$ factors appearing. Thus, for example, to construct the standard model gauge
group in D-brane models 
one starts with the gauge groups $U(3)\times U(2)\times U(1)^2$ which results in
$SU(3)\times SU(2)\times U(1)\times U(1)^3$.  
Typically the breaking of the extra $U(1)$ symmetry will come about by a Higgs mechanism.
Thus if a field has non-vanishing $U(1)$ quantum number, then spontaneous breaking 
which gives a vacuum expectation value  (VEV) to that field will break the extra $U(1)$  
symmetry and  make the extra gauge boson massive.  
This is illustrated by the following example: Under $SU(5)\times U(1)_{\chi}$, 
the $45$ plet of $SO(10)$ has the following decomposition, $45=1(0) + 10(4)+\overline{10}(-4)
+24(0)$. The fields  which can gain vacuum expectation values without violating charge  
or color are  $1(0)$ and $24(0)$, which however has vanishing $U(1)_{\chi}$ quantum numbers.
Thus while the 45 plet breaks $SO(10)$ it does not reduce the rank and the $U(1)_{\chi}$ gauge
boson remains massless. To reduce the rank one needs  a $16$ plet or a $126$ plet of Higgs. 
Thus under  $SU(5)\times U(1)_{\chi}$ the 16 plet of Higgs has the decomposition 
$16=1(-5)+ \bar 5(-3) + 10(-1)$, while the $126$ plet has the decomposition 
$126=1(-10)+ \bar 5(-2)+ 10 (-6) +\overline{15}(6) + 45(2) +\overline{50}(-2)$. 
It is then clear that the VEV growth for $16$ or $126$ will lead to a rank reduction. 
Thus a combination of $45$ and $16$ or  a combination of $45$ and $126$ can reduce 
$SO(10)$ to the standard model 
gauge group.

 More recently it has been shown that one can accomplish the above with a single
irreducible Higgs representation, specifically a $144$ plet representation of $SO(10)$\cite{Babu:2006rp}.
This is so because
under $SU(5)\times U(1)_{\chi}$, the $144$ plet decomposes as follows:
$144=\bar 5(3) +5(7) +10(-1) +15 (-1)+ 24(-5) + 40(-1) +\overline{45}(3)$. In this case
the 24-plet of $SU(5)$ in the above decomposition carries  a $U(1)$ quantum number, and
thus the $SO(10)$  gauge group breaks directly to the SM gauge group after $144$  plet develops
a VEV. 
Further, it may happen that one or more of the extra $U(1)'s$ may remain 
unbroken at the GUT scale, and such breaking may only occur at the electroweak scale.
In this case the mass of the extra gauge boson which we shall generically call a $Z'$ 
will have a mass of the electroweak size. Further, if the extra $U(1)$ is coming from a 
grand unified group the size of the gauge coupling would typically be the size of the
gauge couplings in the electroweak group, i.e., typically of size $g_2$. Thus one generally
expects decay widths for the $Z'$ gauge bosons in such scenarios to be size O(GeV).

The outline of the rest of the paper is as follows:  We first discuss the \st extension of 
the standard model and the electroweak constraints on it\cite{Kors:2004dx,Kors:2005uz}
\cite{Feldman:2006ce,Feldman:2006wb}.
Next  we discuss the \st extension of MSSM\cite{Kors:2004ri}.
This is followed by a discussion of  the
LHC signatures of \st extensions. Of specific interest here is the possibility of a narrow $Z'$. 
Finally we discuss the possibility of dark matter arising from the hidden sector of this 
theory\cite{Cheung:2007ut,Feldman:2007wj}  consisitent with the WMAP\cite{Spergel:2006hy} constraints.
 In this context we also discuss the 
 \st extension to include kinetic energy mixing. Conclusions are given at the end. 
 There are many related works on $U(1)$ extensions (see, e.g., \cite{Carena:2004xs}, and for 
 a review see \cite{Langacker:2008yv})
  as well as works where hidden sectors play a central role\cite{holdom,Kumar:2006gm,Chang:2006fp,Han:2007ae,Gershtein:2008bf,Ahlers:2007qf,Ibarra:2008kn, Georgi:2007ek,Goldberg:2008zz}. 
 Further recent works regarding \st  extensions in the  context
 of the string and D brane models can  be found in \cite{Coriano':2005js,Anastasopoulos:2007qm, Armillis:2007tb} 
and in \cite{Kumar:2007zza,Burgess:2008ri}.

We begin by introducing the  Stueckelberg mechanism\cite{Stueckelberg:1900zz,Cianfrani:2007tu}
for mass growth for $U(1)$ gauge fields.
 In its simplest form  the Stueckelberg mechanism works as  follows: consider  the Lagrangian
 given by 
  \beqn
 L= -\frac{1}{4} F_{\mu\nu} F^{\mu\nu} -\frac{1}{2} (m A_{\mu}+ \partial_{\mu} \sigma)^2 + gA_{\mu} J^{\mu}
 \eeqn
where $\sigma$ is an axionic like field and $J^{\mu}$ is a conserved current.  The Lagrangian is invariant under
the transformation
\beqn
\delta A_{\mu}\to  \partial_{\mu} \epsilon, ~\delta \sigma \to -m \epsilon.
\eeqn
The Stueckelberg mechanism is endemic in extra dimension models, in strings, and in D branes.
Thus consider, for example, the compactification of a 5D theory on $S^1/Z_2$ (see, e.g., \cite{Nath:1999fs}).
As an illustration consider the kinetic energy of a $U(1)$ gauge field  $A_a (a=0,1,2,3,5)$ in 5D, 
\beqn
L_5= -\frac{1}{4} F_{ab}(z)F^{ab}(z), ~a=0,1, 2, 3, 5, 
\eeqn
where $z^a=(x^{\mu}, y)$, $\mu=0,1,2,3$ and we may write $A_a=(A_{\mu}(z),\phi(z))$. 
After compactification on the fifth co-ordinate $y$ 
on a  half circle, one has the Lagrangian in 4D with one mass less mode and an infinite 
number of massive Kaluza-Klein modes
\beqn
L_4=-\frac{1}{4}\sum_{n=0}^{\infty}  F_{\mu\nu}(x)^{(n)}F^{\mu\nu(n)}(x)
-\frac{1}{2}\sum_{n}M_n^2 (A_{\mu}^{(n)} (x) 
+ \frac{1}{M_n}   \partial_{\mu} \phi^{(n)} (x))^2+\cdot\cdot\cdot.
\eeqn
In string models and in D brane models the Stueckelberg phenomenon arises naturally from the 
Green-Schwarz mechanism from a two form field $B_{\mu\nu}$. Thus consider the Lagrangian
with a $U(1)$ gauge field $A_{\mu}$, and a coupling with the two form field $B_{\mu\nu}$ so that
\cite{Kalb:1974yc,Allen:1990gb,Ghilencea:2002da}
\beqn
L_0= -\frac{1}{4} F_{\mu\nu}F^{\mu\nu} -\frac{1}{12} H^{\mu\nu\rho} H_{\mu\nu\rho} 
+\frac{m}{4} \epsilon^{\mu\nu\rho\sigma} F_{\mu\nu} B_{\rho\sigma},  
\eeqn 
where $H_{\mu\nu\rho}$ is the field strength of the two form field $B_{\mu\nu}$ so that 
$H_{\mu\nu\rho}=\partial_{\mu} B_{\nu\rho} + \partial_{\nu} B_{\rho\mu} + \partial_{\rho} B_{\mu\nu}$.
It is convenient to write the last term as \cite{Ghilencea:2002da}
$-\frac{m}{6}  \epsilon^{\mu\nu\rho\sigma} (H_{\mu\nu\rho} A_{\sigma} + \sigma \partial_{\mu} H_{\nu\rho\sigma})$. 
An intergration over $\sigma$ and insertion back gives $L_0$.  Next suppose one
solves for $H^{\mu\nu\rho}$ which gives 
$H^{\mu\nu\rho}= -m  \epsilon^{\mu\nu\rho\sigma} (A_{\sigma} + \partial_{\sigma} \sigma)$. 
An integration on $H^{\mu\nu\rho}$ gives $L_0$ in the form\cite{Ghilencea:2002da}
\beqn
L_1= -\frac{1}{4} F_{\mu\nu}F^{\mu\nu} -\frac{m^2}{2} (A_{\sigma} + \partial_{\sigma} \sigma)^2. 
\eeqn 
Thus we see that the presence of the Green-Schwarz term leads us to mass growth for the $U(1)$ gauge
field by the \st mechanism.

The \st mechanism to give  mass to the vector bosons works whether or not the extra $U(1)_X$ is anomalous.
Thus in general, an anamolous $U(1)$  gives at one loop\cite{Kors:2004iz,Blumenhagen:2006ci}
\beqn
\delta L_{1loop}= \lambda c Tr(G\wedge G), ~\delta A_{\mu} =\partial_{\mu} \lambda. 
\eeqn
This term is cancelled   by variation of the effective tree level term\cite{Kors:2004iz}
\beqn
L_0= \cdot\cdot + mA^{\mu}\partial_{\mu} \sigma + \frac{c\sigma}{m} Tr(G\wedge G),\nonumber\\
\delta L_0= -\lambda c Tr(G\wedge G), ~~~\delta \sigma =-\lambda m,
\eeqn
so that 
\beqn
\delta L_{1loop} +\delta  L_0= 0.
\eeqn
Thus one has the following two cases: 
(i) Anomalous $U(1)$ case, $c\neq 0$, $m\neq 0:$  Here  one cancels the anomaly and at the same time
one has mass growth by the \st mechanism for the $U(1)_X$ gauge vector  boson; 
(ii) Non-anomalous case, $c=0, ~m\neq 0$.  In this case there is no need for anomaly cancellation,
but there is still mass growth for the vector boson by the \st mechanism. Anomalous $U(1)$ in the 
above  context have been discussed by a number of authors (see, e.g.,\cite{anomalous,Anastasopoulos:2007qm,Zagermann:2008gy,DeRydt:2008hw}).

\noindent
{\em \st vs the Higgs  mechanism:}  Next we explore the connection between the \st mechanism and the
Higgs mechanism\cite{Allen:1990gb}.
Consider a $U(1)$ gauge  theory with a Higgs  potential  so that 
\beqn
L_0=  -\frac{1}{4} F_{\mu\nu}F^{\mu\nu} -(D^{\mu}\phi)^{\dagger} D_{\mu} \phi  -V(\phi) +L_{gf}, \nonumber\\
V(\phi) = \mu^2 (\phi^{\dagger} \phi)  + \lambda (\phi^{\dagger} \phi)^2.  
\eeqn 
With $\mu^2<0$, and $\lambda >0$  a spontaneous breaking of the $U(1)$ gauge symmetry occurs, and one
has 
\beqn
\phi =\frac{1}{\sqrt {2}} (\rho + v)e^{ia/v},  v=\sqrt{ -\mu^2/\lambda}.
\eeqn
 In the limit $(-\mu^2, \lambda)\to \infty$ with $M=ev$ fixed, $\rho$ becomes infinitely heavy 
 and decouples from the rest of the system, and the residual Lagrangian is given by\cite{Allen:1990gb}
 \beqn
L= -\frac{1}{4} F_{\mu\nu}F^{\mu\nu} -\frac{1}{2} M^2 (A_{\mu} -\frac{1}{M} \partial_{\mu} a)^2
+ L_{gf}.
\eeqn
Thus we see that in the limit considered above the Higgs mechanism leads to the \st mechanism 
in a very direct way.\\

\noindent
{\em \st extension of the Standard Model:}
Although the \st mechanism has been around for a long
time, a successful incorporation of this mechanism into particle physics  was made  only 
recently\cite{Kors:2004dx,Kors:2004ri,Kors:2004iz}
and its phenomenological implications investigated\cite{Kors:2005uz,Feldman:2006ce,Feldman:2006wb,Feldman:2006wd,Feldman:2007wj,Feldman:2007nf}.
Specifically one can extend the standard model  by considering the gauge group
$SU(3)_C\times SU(2)_L\times U(1)_Y\times U(1)_X$. We will assume that the visible sector matter
fields (quarks, leptons, Higgs) are  neutral under the gauge group $U(1)_X$, while the hidden sector
fields are neutral under the standard model gauge group. Let  the gauge fields  for  $U(1)_Y$ and 
$U(1)_X$ be $B_{\mu}$ and $C_{\mu}$.  We assume that there  exists a connector sector which 
carries dual quantum numbers, i.e., quantum numbers of the gauge group $U(1)_Y$ and of $U(1)_X$.
Specifically we  assume that the axionic field $\sigma$ is indeed  this connector field. Then we append to the
SM Lagrangian the following \st extension 
\beqn
L_{St} = -\frac{1}{4} C_{\mu\nu} C^{\mu\nu} + g_X C_{\mu} J^{\mu}_X 
-\frac{1}{2} (\partial_{\mu} \sigma + M_1 C_{\mu} + M_2 B_{\mu})^2,
\eeqn
where $J_X^{\mu}$ is a conserved current arising from the hidden sector.
The above  Lagrangian is  invariant under the $U(1)$ transformations
\beqn
\delta_Y B_{\mu} =\partial_{\mu} \lambda_Y, ~~ \delta_Y C_{\mu}=0,  
~~\delta_Y \sigma = -M_2 \lambda_Y, 
\eeqn
and under the $U(1)_X$ transformations 
\beqn
\delta_X C_{\mu} =\partial_{\mu} \lambda_X, ~~\delta_X B_{\mu}=0,  
~~\delta_X \sigma = -M_1 \lambda_X. 
\eeqn
The total Lagrangian is of course $L_{SM} + L_{St}$.  After spontaneous breaking of the
electroweak symmetry, the above Lagrangian gives  a  mass$^2$ matrix of the form 
\beqn
M^2_{ab} = 
\left(\matrix{ M_1^2  &  M_1M_2  &  0\cr
M_1M_2 & M_2^2 + \frac{1}{4} g_Y^2 v^2 & - \frac{1}{4}g_Yg_2 v^2
\cr 0 & -\frac{1}{4}g_Yg_2 v^2 & \frac{1}{4}g_2^2 v^2
}\right)\ ,
\label{vm2}
\eeqn
where we use a basis $(V_\mu^{\rm T})_a = ( C_{\mu}, B_{\mu}, A_{\mu}^3)_a$ and use the 
standard form  $-\frac12 V_{a\mu} M^2_{ab} V^\mu_b$ for the mass term in the Lagrangian. 
In Eq.(\ref{vm2}), $g_2$ and $g_Y$ are the $SU(2)_L\times U(1)_Y$ gauge coupling constants
and $v=2M_{\rm W}/g_2=(\sqrt 2 G_F)^{-\frac{1}{2}}$,  where
$M_{\rm W}$ is the mass of the W boson, and $G_F$ is the Fermi
constant.


 The eigen modes of the mass matrix of 
the vector bosons give a mass less mode which is a photon $\gamma$ and two massive modes
which are $Z$ and $Z'$. The definition of the electric charge is modified so that
\beqn
\frac{1}{e^2} =\frac{1}{g_2^2} + \frac{ 1+\epsilon^2}{g_Y^2},
\eeqn
where $\epsilon =M_2/M_1$. Fits to the precision electroweak data require that $\epsilon$ be very
small, i.e., $\epsilon\leq .06$\cite{Feldman:2006ce}. The smallness of $\epsilon$ leads to the existence of a very narrow 
resonance if there is no matter in the hidden sector, or if the $Z'$ is forbidden kinematically 
 to decay into the hidden sector matter. This can happen if, for example, $M_{Z'}< 2M_{hid}$
where $M_{hid}$ is the mass of the hidden sector fermion. Further, it also follows that if there is matter in
the hidden sector, such matter will couple to the photon with a milli charge strength, i.e., with strength
$Q_{\epsilon}=\epsilon e$. Detailed fits to the LEP I data show that one can satisfy the LEP I constraints to 
essentially the same level of confidence as fits to the standard model. \\

\noindent
{\em LEP II constraints:} These arise from the  constraints on the contact interaction\cite{:2003ih}
\beqn
L_C\sim \frac{g^2\eta_{sign}}{(1+\delta) \Lambda^2} 
\sum_{i,j=L,R}
\bar e_i \gamma^{\mu} e_i \bar f_j \gamma_{\mu} f_j.
\eeqn
where $\delta =0$ for $f\ne  e$, and  $\delta =1$ for $f=  e$,
 $\eta_{ij}$ gives the relative contribution of the different chiralities, and $\eta_{sign}$
tells us if the contribution is constructive or destructive relative  to the SM contribution. 
The most stringent constraints arise from $\Lambda_{VV}$ for which LEP II gives 
\beqn
\Lambda_{VV}> 21.7 {\rm GeV},
\eeqn
while the \st extended model gives\cite{Feldman:2006wb}
\beqn
\Lambda_{VV} =\frac{M_{Z'}}{M_Z} \left(\frac{4\pi}{\sqrt 2 G_F v_e^{'2}}\right)^{1/2}.
\eeqn
For the \st extension the LEP II constraints are automatically  satisfied 
once one satisfies the LEP I constraints at the 1\% level. \\

The \st extension differs from the previous models in that the photon field is now a linear combination
of three gauge fields, $A_{\mu}^3, B_{\mu}, C_{\mu}$ so that
\beqn
A_{\mu}^{\gamma} =-c_{\theta}s_{\phi} C_{\mu} + c_{\theta}c_{\phi} B_{\mu} +s_{\theta} A_{\mu}^3. 
\eeqn
This is to be compared with the form of  $A_{\mu}^{\gamma}$ in the standard model  where one has 
$A_{\mu}^{\gamma} = c_{\theta} B_{\mu} +s_{\theta} A_{\mu}^3$, so it is a combination of only the fields
$B_{\mu}$ and $A_{\mu}^3$. It is precisely because $A_{\mu}^{\gamma}$ has a small dependence on
the field $C_{\mu}$ that the  photon is able to couple to the hidden sector matter with milli charge
strength. The model predicts a very narrow $Z'$ resonance
in absence of its decay into the hidden sector matter which is in contrast to rather broad resonances
arising from Kaluza-Klein excitation of a compact extra dimension\cite{Antoniadis:1999bq,Nath:1999mw}.
Electroweak tests of the \st extension of the standard model can be found 
in \cite{Kors:2005uz,Feldman:2006ce}.\\

\noindent
{\em Stueckelberg extension of the minimal supersymmetric standard model:}
We consider now the Stueckelberg extension of the minimal supersymmetric standard model(MSSM),
and to this end we introduce  vector super fields for the $U(1)_Y$ and the $U(1)_X$ gauge groups which
we label as $B$ and $C$ respectively.  Additionally we introduce the chiral multiplets $S$ and $\bar S$ 
which contain the axionic field and we choose for the \st  Lagrangian the form 
\beqn
\int d\theta^2  d \bar\theta^2 (M_1C +M_2 B + S+\bar S)^2. 
\eeqn  
The above Lagrangian is invariant under the following $U(1)_Y$ and $U(1)_X$ transformation 
\beqn
\delta_Y  B= \Lambda_Y +\bar \Lambda_Y, ~ \delta_Y S= -M_2 \Lambda_Y,\nonumber\\
\delta_X  C= \Lambda_X +\bar \Lambda_X, ~ \delta_X S= -M_1 \Lambda_X.
\eeqn 
Additionally, of course, we have the gauge kinetic energy terms for gauge multiplet $C$ while the kinetic
energy terms for the $B$ multiplet are contained in the MSSM Lagrangian.  In the Wess-Zumino gauge
the components of the gauge multiplet $C$ are $(C_{\mu}, \lambda_ C, \bar \lambda_C, D_C)$, 
and similarly  for the gauge multiplet $B$, while the chiral multiplet $S$ has the components
$S= (\rho + i\sigma, \chi, F_S)$ and similarly for $\bar S$.  It is then possible to form two 
additional Majorana spinors beyond those in MSSM. Here one has
\beqn
\psi_S=
 \left(
\begin{array}{c}
\chi_{\alpha}\\ 
 \bar\chi^{\dot\alpha}
\end{array}\right),  
~~\lambda_X=
 \left(
\begin{array}{c}
\lambda_{C\alpha}\\ 
 \bar\lambda_C^{\dot\alpha}
\end{array}\right). 
\eeqn
Including the above two, one has six Majorana  basis states 
\beqn
\psi_S, \lambda_X, \lambda_Y, \lambda_3, \tilde h_1, \tilde h_2,
\eeqn
where  $\lambda_Y, \lambda_3, \tilde h_1, \tilde h_2$ are the four familiar gaugino and higgsino
states of MSSM.  As in SUGRA models, there is mixing among the six neutral states which gives 
rise to the six neutralino states in the mass diagonal basis. We can label these states as
\beqn
\xi_1^0, \xi_2^0, \chi_1^0, \chi_2^0, \chi_3^0, \chi_4^0,
\eeqn
where $\chi_i^0 ~(i=1-4)$ are the familiar four neutralino states in MSSM, and $\xi_1^0$ and
$\xi_2^0$ are the two additional neutral states that arise from the \st sector.  
An interesting situation arises when the LSP of the entire system is in the \st sector, i.e., 
it is $\xi_1$ which we assume is the lighter of $\xi_{i} (i=1,2)$. We will discuss this possibility 
in the context of dark matter later.\\

\noindent
{\em Higgs sector in \st extension of MSSM:} The Higgs sector is affected in  the \st extension of
MSSM. This is so because one has in the \st extension of MSSM the field $\rho$  which couples 
with the CP even MSSM higgs fields $H^0, h^0$  giving a $3\times 3$ mass matrix. We display 
this mass matrix below\cite{Kors:2004iz}
\beqn
(M^2_{H^0}) = 
\left(\matrix{ (M_Z^2c^2_{\beta} +m_A^2 s^2_{\beta}) 
 &  -(M_Z^2+m_A^2)s_{\beta}c_{\beta}  & -s_{\theta}c_{\beta} M_Z M_2 \cr
 -(M_Z^2+m_A^2)s_{\beta}c_{\beta}& (M_Z^2s^2_{\beta} +m_A^2 c^2_{\beta})& 
s_{\theta}s_{\beta} M_Z M_2  \cr 
-s_{\theta}c_{\beta} M_Z M_2  & s_{\theta}s_{\beta} M_Z M_2 & 
m_{\rho}^2}\right)\ ,
\eeqn
where $s_{\beta} (c_{\beta})=\sin\beta (\cos\beta)$, 
$s_{\theta}=\sin\theta_W$ where $\theta_W$ is the weak angle. 
The above matrix in the neutral Higgs sector has the eigen states
$H_1^0, H_2^0, H_3^0$. One may choose these so that 
$H_1^0, H_2^0, H_3^0$ $\rightarrow$ $H^0, h^0, \rho$ for the case when
the coupling to the Stueckelberg sector vanishes, i.e., $M_2=0$,
where $h^0$ is the light neutral Higgs of MSSM and $H^0$ is the heavy 
neutral Higgs of MSSM.\\

\noindent
{\em The \st mechanism and the Higgs mechanism in MSSM:}  In the non-supersymmetric case it was
shown that the Higgs mechanism reduces to the \st mechanism\cite{Stueckelberg:1900zz}
in an appropriate limit with a 
constraint on the mass$^2$  and coupling constant parameters of the Higgs potential.  
In the supersymmetric case, it turns out that one needs  Fayet-Iliopoulos D-terms  to accomplish
this reduction\cite{Feldman:2006wd}.  Thus we begin by adding Fayet-Iliopoulos D terms to the scalar potential
\beqn
L_{FI}= \xi_X D_C + \xi_Y D_Y
\eeqn
 Including these the D-part of the scalar potential  has the form
 \beqn
 V_D= \frac{g_X^2}{2}  (Q_X|\phi^+|^2 - Q_X|\phi^-|^2+\xi_X)^2 
  +  \frac{g_Y^2}{2} (Y_{\phi}|\phi^+|^2 - Y_{\phi}|\phi^-|^2+\xi_Y)^2  
  \eeqn
 Minimization of the potential gives $\langle \phi^+\rangle =0$, $\langle \phi^-\rangle \neq 0$  and 
 \beqn
 M_1 =\sqrt 2 g_X Q_X \langle \phi^-\rangle,    M_2 =\sqrt 2 g_Y Y_{\phi} \langle \phi^-\rangle,  
  \eeqn
The limit $g_X Q_X \to 0, g_Y Y_{\phi} \to 0$ with $M_1, M_2$ fixed reduces the Higgs mechanism
above to the \st form\cite{Feldman:2006wd}. \\

\noindent
{\em Kinetic mixing:} The kinetic energy of two $U(1)'s$ will mix if there are fields even with GUT 
size mass which carry dual quantum numbers\cite{holdom}. 
Thus consider two $U(1)$ gauge fields $A_1^{\mu}, A_2^{\mu}$  with the interaction 
\beqn
L_{km}= -\frac{1}{4} F_{1\mu\nu} F_1^{\mu\nu} -\frac{1}{4} F_{2\mu\nu} F_2^{\mu\nu}
-\frac{\delta}{2} F_{1\mu\nu} F_2^{\mu\nu} + J_{\mu}' A_1^{\mu} + J_{\mu} A_2^{\mu}.
\eeqn
where $J_{\mu}$ is the source arising from the visible sector matter fields and $J_{\mu}'$ is the source
containing fields in the hidden sector. We can go to the diagonal basis with the transformation 
\beqn
\left(\begin{array}{c}
 A_1^{\mu}\\ 
 A_2^{\mu}
\end{array}\right)
 \to K_0
 \left(\begin{array}{c}
 A^{'\mu} \\ 
A^{\mu}
\end{array}\right), ~~K_0= 
\left(\begin{array}{c}
 \frac{1}{\sqrt{1-\delta^2}} ~~0\\ 
  \frac{-\delta}{\sqrt{1-\delta^2}} ~~1\end{array}\right).
\eeqn
However, there is a degree of arbitrariness in the choice  of the matrix that diagonalizes the kinetic energy 
matrix. Specifically one may take $K$ instead of $K_0$ as the diagonalizing matrix where 
\beqn
K=K_0 R, ~~ R= \left(\begin{array}{c}
 \cos\theta  ~-\sin\theta \\ 
\sin\theta ~~ \cos\theta\end{array}\right).
\eeqn
A convenient choice of $\theta$ is 
\beqn
\theta = arctan [\delta/\sqrt{1-\delta^2}]
\eeqn
which gives the asymmetric solution 
\beqn
L_1^{K}=
 A^{\mu} \left[ \frac{1}{\sqrt{1-\delta^2}} J_{\mu} -\frac{\delta}{\sqrt{1-\delta^2}} J_{\mu}'\right] 
+A^{\mu '} J_{\mu}'  
\eeqn
We identify $A^{\mu}$ with the photon field which has interactions with the visible sector source $J_{\mu}$ 
with normal strength and also an interaction with the hidden sector source $J_{\mu}'$ with a strength
which is proportional to $\delta$ and thus this interaction is milli charge size. The field $A_{\mu}'$ is 
another massless vector field which  interacts with the hidden sector source $J_{\mu}'$ and has no
interactions with the visible sector current $J_{\mu}$.\\ 

Next let us consider the \st mechanism with kinetic energy mixing.  Here in addition to the kinetic mixing
one also has mass mixings so that 
\beqn
L_{St}^m= -\frac{1}{2} M_1^2 A_{1\mu} A_1^{\mu} -\frac{1}{2} M_2^2 A_{2\mu} A_2^{\mu}
-M_1M_2 A_{1\mu} A_2^{\mu}.
\eeqn
Diagonalization of the mass matrix fixes $\theta$ so that 
\beqn
\theta= arctan \left( \frac{\epsilon \sqrt{1-\delta^2}}{1-\epsilon \delta}\right).
\eeqn
With the above one finds the following interactions in the basis where both the mass and
the kinetic energy terms are diagonal 
\beqn
L_{St}^{int}  = \frac{1}{\sqrt{1-2 \epsilon \delta +\epsilon^2}} \left(\frac{\epsilon -\delta}{\sqrt{1-\delta^2}} J_{\mu}
+\frac{1-\epsilon\delta}{\sqrt{1-\delta^2}} J_{\mu}' \right) A_M^{\mu}\nonumber\\
+\frac{1}{\sqrt{1-2 \epsilon \delta +\epsilon^2}} \left( J_{\mu}
-\epsilon   J_{\mu}' \right) A_{\gamma}^{\mu}. 
\eeqn 
Here $A_{\gamma}^{\mu}$ is the photonic field and $A_{M}^{\mu}$ is the massive vector boson field. 
We note that the coupling of the photon with the hidden sector is controlled by $\epsilon$ and 
vanishes when $\epsilon$ vanishes. This is in contrast to the case of the pure kinetic mixing where
the coupling of the photon with the hidden sector is controlled by $\delta$. One can carry out 
a similar extension of the \st extension of the standard modelgauge group to 
$SU(3)_C\times SU(2)_L\times U(1)_Y\times U(1)_X$ with both kinetic mixing and mass mixing for the 
two $U(1)'s$. In this case one finds that the electric charge is modified so that 
\beqn
\frac{1}{e^2} = \frac{1}{g_2^2} + \frac{1-2\epsilon\delta + \epsilon^2}{g_Y^2}.
\eeqn
An interesting result arises in that in the absence of the hidden  sector, the weak sector of the model
depends only on the combination $\bar \epsilon=(\epsilon-\delta)/\sqrt{1-\delta^2}$. In this case the 
precision LEP data constrains $\bar \epsilon$\cite{Feldman:2007wj}.  
 For a discussion of other approaches to $U(1)$ probes of the hidden sector see 
 \cite{Han:2007ae,Gershtein:2008bf,Kumar:2007zza}.\\

\noindent
{\em Milli-weak and hidden sector dark matter in \st extensions:}
The \st extension gives rise to two new candidates for dark matter. One of these is milli weak or extra weak
dark matter arising from the \st extension of MSSM. The other is  dark matter which arises
from the hidden sector to which $U(1)_X$ couples. We discuss each
of these cases in some detail below.\\

\noindent
{\em Milli weak dark matter:}  In the \st extension of  MSSM one finds  6 neutralino 
states four of which are the usual states from the MSSM sector while the remaining two arise from the
\st sector. Here there are two distinct possibilities\cite{Kors:2004ri}.
The first one corresponds to the case when
$\chi_1^0$ is the LSP. 
In this case 
 the analysis of dark matter and of the supersymmetric signatures would
exactly be the same as for the case of the SUGRA models except for some minor corrections arising  
from the mixings of the \st and the MSSM sector. The second possibility corresponds to the case
when $\xi_1^0$ is the LSP. In this case the LSP is milli weak or extra weakly interacting.  The LSP
of the MSSM sector would decay into $\xi_1^0$ inside the detector, and thus one still has missing 
energy signals.
Naively one might expect that 
 because of the extra weak interactions of $\xi_1^0$ it might be difficult to 
 achieve an efficient annihilation of the $\xi_1^0$'s. However, this is not the case when one takes
into account the coannihilations. It is then possible to satisfy the relic density constraints consistent
with WMAP data\cite{Feldman:2006wd,Fucito:2008ai}.
 For other possible candidates arising from $U(1)$ extensions 
see\cite{Barger:2004bz,Hur:2007ur}.\\

\noindent
{\em Hidden sector dark matter}\cite{Cheung:2007ut,Feldman:2007wj}:
Hidden sector fermions could be candidates for dark matter. Specifically let us assume
that on has massive Dirac fermions with mass  $m_D$  in the hidden sector. 
These would be milli charged (for other recent works with milli charged particles see
 \cite{Huh:2007zw} 
 
 Detailed  analyses then show that one can satisfy the WMAP relic 
density constraints\cite{Cheung:2007ut,Feldman:2007wj}. 
There are two clear regimes for the relic density constraints to be satisfied.
The first one is when $ M_{Z'}> 2m_{D}$ in which case the $Z'$ width will be normal size because 
$Z'$ can decay with normal strength into the Dirac fermions of the hidden sector. However, the $Z'$
coupling with the SM quarks and leptons is small because it is suppressed by a factor $\epsilon^2$.
Consequently in this scenario it is difficult to detect the $Z'$ by Drell-Yan at the Large Hadron Collider.  
The second possibility corresponds to the case when $M_{Z'} <2m_{D}$. 
Here it is still possible to satisfy the relic density constraints by a proper thermal averaging over the $Z'$ pole. 
However, in this case  the $Z'$ cannot decay on shell  into the Dirac fermions of the hidden sector
but it can do so into the quarks and leptons of SM. Thus one can detect the $Z'$ by the Drell -Yan 
at the Large Hadron Collider. 
The Dirac fermions in the hidden sector can explain the positron excess seen by  
 the anti-matter satellite probe PAMELA\cite{Adriani:2008zr}. An analysis shows that 
   such an excess can 
be understood from the annihilation of Dirac fermions in the hidden sector\cite{Feldman:2008xs}.
An important issue emphasized in the analysis of \cite{Feldman:2008xs} is the enhancement or
boost  of the ratio $\langle \sigma v \rangle_{Halo}/\langle \sigma v \rangle_{Freezeout}$ 
near a pole due to the fact that the temperature in the halo is 
 much smaller compared to the temperature at the freezeout. 
The above phenomenon come about when twice the mass of the annihilating particle  is  close to the resonance
mass in the channel in which they annihilate. For the annihilation of the Dirac particles considered in 
\cite{Feldman:2008xs} the resonance is the \st $Z'$ pole. 
In this case the ratio is very sensitive to the temperature, and the velocity averaging in the halo gives larger
results due to a much smaller temperature in the halo relative to the velocity averaging at the freezeout 
where the temperature is much larger $T \sim m_D/(20-30)$. Because of the above enhancement much smaller additional boost factors are needed to fit the positron data\cite{Feldman:2008xs}.
For other works on the \st mechanism and applications see \cite{Zhang:2008xq,Burgess,Ahlers:2007qf}.
and for other works on dark matter from hidden sector see\cite{Huh:2007zw,Chun:2008by}.\\

\noindent
{\em Conclusion:}
Extra $U(1)$'s arise in a wide variety of GUT, string and D brane models, and the \st extensions
of SM  provide a natural framework to incorporate them. The \st mechanism is an alternative to 
the Higgs mechanism for the breaking of the $U(1)$ gauge symmetry. It has the advantage over the
Higgs mechanism in that one does not need to construct a scalar potential or  generate spontaneous 
breaking to  give mass to the gauge vector boson.  The \st extension leads to new phenomena 
testable at colliders, specfically a sharp $Z'$ resonance with a width in the MeV to sub GeV range
in the absence of decay of the $Z'$ into hidden sector matter. 
Two new types of dark matter emerge in the \st extension. The first is a milli weak dark matter 
candidate whose interactions with matter are weaker than the weak interactions of the weakly
interacting massive particles (WIMPS). The second new candidate for dark matter that 
emerges is the milli charged dark matter which arises solely from the hidden sector. 
It is found that both the milli weak and the milli charged dark matter candidates can generate the 
right amount of dark matter in the universe consistent with the WMAP relic density constraints. 
Since the \st mechanism arises quite naturally in string and D brane models, a confirmation the
\st mechanism experimentally would lend support to the existence of a new  regime of physics - 
perhaps string theory. \\

\noindent
{\em Acknowledgments:}
The author thanks Professor Gerard t'Hooft and Professor Antonio Zichichi for 
invitation to lecture at the 46th Course in Subnuclear Physics at Erice, and  
Professor Zichichi for the hospitality extended  him during the period of the
course. The new work reported here is based on collaborative work with
K.S. Babu, Ilia Gogoladze, Daniel Feldman, Boris Kors, Zuowei Liu, and
Raza Syed.    This research is  supported in part by NSF grant PHY-0757959.

\end{document}